\documentclass[11pt,a4paper]{article}

\usepackage[utf8]{inputenc}
\usepackage[T1]{fontenc}
\usepackage{amsmath,amssymb}
\usepackage{bm}
\usepackage{graphicx}
\usepackage{booktabs}
\usepackage{hyperref}
\usepackage{url}
\usepackage[numbers,sort&compress]{natbib}
\usepackage{authblk}
\usepackage[margin=2.5cm]{geometry}
\usepackage{caption}
\usepackage{float}
\usepackage{enumitem}
\usepackage{xcolor}

\hypersetup{
  colorlinks=true,
  linkcolor=blue,
  citecolor=blue,
  urlcolor=blue
}

\newcommand{\Us}{U^{*}}

\newcommand{\vect}[1]{\bm{#1}}
\newcommand{\doi}[1]{\href{https://doi.org/#1}{doi:#1}}
\newcommand{\orcid}[1]{\textsuperscript{\href{https://orcid.org/#1}{\scriptsize ORCID}}}

\title{\textbf{Piezoelectric Energy Harvesting from a Pitch-Plunge-Flap Aerofoil
in Compressible Flow, the Euler Full-Order Model, Strip Theory and the Reduced
Models Compared}}

\author[1]{Nikolaos~D.~Tantaroudas\orcid{0000-0001-6727-1014}\thanks{Corresponding
author: \texttt{nikolaos.tantaroudas@iccs.gr}.}}
\author[2]{Ilias~Karachalios\orcid{0009-0000-8581-8335}}
\author[3]{Andrew~J.~McCracken}

\affil[1]{Institute of Communications and Computer Systems (ICCS), National
Technical University of Athens, Iroon Polytechniou 9, 15773 Zografou, Greece}
\affil[2]{Department of Environmental Sciences, University of Thessaly, Gaiopolis
Campus, Larissa-Trikala Ring Road, 41500 Larissa, Greece}
\affil[3]{DASKALOS APPS, 183 Rue de l'Abb\'{e} Griffon, 01960 P\'{e}ronnas,
France}

\date{}

\begin{document}

\maketitle

\begin{abstract}
A piezoelectric transducer embedded in a pitch-plunge aerofoil with a finite-mass
trailing-edge flap converts the limit-cycle oscillation that follows flutter into electrical
power. The design findings for such a harvester have so far been obtained with incompressible
strip theory, whose aerodynamic state is a handful of lag variables. Here the same section, the
same transducer and the same test cases are coupled instead to the two-dimensional compressible
Euler equations, giving a full-order model of twelve thousand states, and the two
aerodynamic models are placed side by side on the flutter boundary, on the limit cycle and on the
harvested power, with the strip-theory results compared against higher fidelity aerodynamic modelling provided by CFD. The degree of freedom carrying the transducer
remains the first-order design variable, and the ranking of the mountings and the position of the
optimum in the coupling and load-resistance plane are unchanged,
while strip theory is found to under-predict the harvested power of the compressible section and
to overstate the effect of the transducer on stability where the electrical time constant meets
the flutter frequency. A nonlinear reduced model of four states, built on eigenvectors of the
coupled Jacobian at the velocity of interest, reproduces the frequency of the cycle, its pitch
amplitude, the transducer voltage and the mean power, and misstates the proportion of plunge to
pitch in the orbit and its phase. That error is shown to be independent of the amplitude of the
cycle and to survive enlargement of the basis, which places it in the retained subspace rather
than in the order of the expansion, and suggests that such models should be judged on the shape of the
orbit rather than on a single amplitude.
\end{abstract}

\section{Introduction}
\label{sec:intro}

A companion paper~\citep{tantaroudas2026harvest}, arXiv:2609.12788, studied piezoelectric energy harvesting from
the post-flutter limit cycle of a three-degree-of-freedom aerofoil, a pitch-plunge section with
a finite-mass trailing-edge flap, using unsteady strip theory for the aerodynamics. It was found
that the degree of freedom carrying the transducer sets both the sign of the shift of the
flutter boundary and the harvested power, and that the error of a four-state nonlinear reduced
model lies in how the reduced operator is carried with flow speed and in the harvested voltage
rather than in the size of the basis. Strip theory is incompressible, linear in the flow and
carries the unsteady lag in a handful of fitted exponentials. The questions left open are
whether those findings survive a compressible flow, in which the aerodynamic states are a whole
field rather than eight lag variables, and whether a reduced model of the same kind can be
built on such a flow at all, since the modes of a flow solver's Jacobian are thousands of
convected and acoustic modes among which the aeroelastic ones must be found.

The present paper repeats the test cases of the companion paper with the aerodynamics replaced
by the two-dimensional compressible Euler equations, coupled to the same structure and the same
transducer, and puts the two aerodynamic models side by side on every result. The full-order
model coupled with CFD has about twelve thousand states. The linear flutter boundary against load
resistance is computed for the transducer on each of the three degrees of freedom, the limit
cycle and its harvested power are marched with the full model across the reduced velocity,
the nonlinear reduced model is built on the eigenmodes of the coupled Jacobian and compared
with the full model in the same quantities the companion paper reports, and the mode that
carries the harvested voltage in the flow model is identified.

\section{Electro-aeroelastic model}
\label{sec:model}

\subsection{Structure and transducer}

The three degree of freedom aerofoil section plunges by $\xi = h/b$, positive downward,
pitches by $\alpha$ about the elastic axis, positive nose-up, and rotates its flap by $\delta$
about the hinge, positive trailing-edge-down. Lengths are scaled by the semi-chord $b$, time
as $\tau = Ut/b$ and flow speed as the reduced velocity $\Us = U/(b\omega_\alpha)$, and primes
denote $\mathrm{d}/\mathrm{d}\tau$. With $\vect{q} = (\xi,\alpha,\delta)^{\mathsf T}$ the
equations of motion are
\begin{equation}
\vect{M}\vect{q}'' + \vect{C}\vect{q}' + \vect{K}\vect{q}
+ \vect{f}_{\mathrm{nl}}(\vect{q}) + \kappa\,\vect{e}_{j}\,v = \vect{f}_{a},
\label{eq:struct}
\end{equation}
where the $3\times3$ matrices carry the structural inertia, damping and stiffness, the
restoring forces are polynomial with only the cubic pitch spring
$(1/\Us)^2\beta_\alpha\alpha^3$ nonzero, $\vect{f}_a$ is the aerodynamic generalised force and
the last term on the left is the back-coupling of the transducer, $\vect{e}_j$ selecting the
degree of freedom that carries it. The transducer is a lumped element with coupling $\theta$
and capacitance $C_p$ shunted by a resistance $R$. With its voltage scaled as
$v = C_pV/(\theta b)$ the circuit is one first-order equation driven by the velocity of the
mounted degree of freedom,
\begin{equation}
v' = q_j' - \lambda v, \qquad \lambda = \frac{1}{\Us r},
\label{eq:circuit}
\end{equation}
and the harvester is set by two dimensionless groups, the coupling
$\chi = \theta^2/(k_jC_p)$ and the load resistance $r = RC_p\omega_\alpha$, with
$\kappa = \chi k_j$ and $k_j$ the linear stiffness of the mounted degree of freedom,
$(\omega/\Us)^2$ for plunge, $1/\Us{}^2$ for pitch and $(\omega_\delta/\Us)^2$ for flap. The
instantaneous power in the load is $\bar p = \chi v^2/r$ and the mean power
$\langle\bar p\rangle$ is its average over a whole number of cycles. Powers from different
mountings are referred to the common reference $P/(mb^2\omega_\alpha^3)$ by the factors
$\omega^2$, $r_\alpha^2$ and $r_\delta^2\omega_\delta^2$. The parameters are those also used
on the companion paper, $\mu = 100$, $\omega = 0.2$, $\omega_\delta = 3.5$, $a_h = -0.5$,
$x_\alpha = 0.25$, $r_\alpha = 0.5$, $c = 0.5$, $r_\delta = 0.0791$, $x_\delta = 0.0125$, no
structural damping, and $\beta_\alpha = 100$ wherever a limit cycle is marched, which holds
the pitch amplitude within a few degrees. The elastic axis at $a_h = -0.5$ lies at the quarter
chord and the hinge at three quarters.

\subsection{Strip-theory aerodynamics}

In the companion paper $\vect{f}_a$ is the incompressible unsteady strip theory of the
pitch-plunge-flap section of Lee et al.~\citep{lee1999} as implemented
in~\citep{tantaroudas2015phd}, with the non-circulatory and quasi-steady terms folded into
$\vect{M}$, $\vect{C}$ and $\vect{K}$ and the circulatory lag carried by six Wagner states,
two exponents per degree of freedom, and two K\"ussner gust states that are inert here. The
coupled system has fifteen states with the transducer.

\subsection{Compressible Euler aerodynamics}

The flow model replaces $\vect{f}_a$ by the integrated surface pressure of the two-dimensional
compressible Euler equations about a NACA~0012 section at a fixed Mach number. The equations
are discretised in method-of-lines form on a structured O-mesh with a cell-centred finite
volume scheme, central fluxes with scalar artificial dissipation of the Jameson type, and a
characteristic far-field condition, so that the flow state $\vect{w}$ of $4n_c$ conserved
variables on $n_c$ cells satisfies
\begin{equation}
\vect{w}' = \vect{R}(\vect{w},\vect{q},\vect{q}'),
\label{eq:flow}
\end{equation}
with the structural motion entering through the mesh, which pitches and plunges rigidly with
the section and deforms smoothly near the flap. Two departures from a production scheme are
deliberate. The pressure sensor of the dissipation is squared rather than taken in absolute
value and its coefficients are blended by a smooth rational function, and there are no
limiters, so that $\vect{R}$ is a smooth function of its arguments. The reduced model
differences it to third order, which a merely continuous residual would not survive. The mesh
has $96\times32$ cells with the far field at 20 chords, the mesh on which the gust response
of the same section was found converged, and it is checked against $64\times24$ and
$128\times48$ on the flutter boundary. Collecting
$\vect{x} = (\vect{q},\vect{q}',\vect{w},v)$ gives the coupled full-order model
\begin{equation}
\vect{x}' = \vect{F}(\vect{x};\Us),
\qquad \vect{x}\in\mathbb{R}^{n}, \quad n = 6 + 4n_c + 1,
\label{eq:fom}
\end{equation}
with $n = 11\,911$ on the paper mesh. Its Jacobian is assembled analytically block by block,
the flow block by differentiating the flux sweep, the coupling blocks by the derivatives of
the loads with respect to the flow state and of the residual with respect to the mesh motion,
and the transducer as the same row and column as in strip theory, and it is held sparse.
The equilibrium is the steady flow about the undeflected section, computed once per Mach
number and velocity. Time responses are marched by the second-order backward difference
scheme with a fixed step of $0.2$ in $\tau$, about $370$ steps per cycle, each step solved by
Newton's method on the sparse Jacobian, which is factorised once and refreshed only when the
iteration stalls. The strip-theory responses of the companion paper were marched by Heun's
scheme at a step of $0.05$.

The Mach number of the comparison is $0.10$, the lowest at which the scheme runs reliably
for this section and the closest to the incompressible reference, and the Mach trend is shown
where it matters. The section itself differs from the flat plate of strip theory by its
thickness, so a residual difference of a few percent between the two models is expected even
in the incompressible limit.

\section{Reduced models of the coupled system}
\label{sec:rom}

The ROM method used in~\citep{daronch2013gust,daronch2013control},
is applied to Eq.~\eqref{eq:fom}. The residual is expanded about the equilibrium,
\begin{equation}
\vect{F}(\vect{x}) = \vect{A}\vect{x}
+ \tfrac{1}{2}\vect{B}(\vect{x},\vect{x})
+ \tfrac{1}{6}\vect{C}(\vect{x},\vect{x},\vect{x})
+ \mathcal{O}(\|\vect{x}\|^{4}),
\label{eq:taylor}
\end{equation}
and projected onto $m$ eigenvectors of $\vect{A}$ with their biorthonormal left eigenvectors,
$\vect{\Psi}^{\mathsf H}\vect{\Phi} = \vect{I}$, so that
$\vect{x} = 2\,\mathrm{Re}(\vect{\Phi}\vect{z})$ for complex pairs and the reduced system in
the coordinates $\vect{z}$ is
\begin{equation}
\vect{z}' = \vect{\Lambda}\vect{z}
+ \tfrac{1}{2}\vect{B}_{r}(\vect{z},\vect{z})
+ \tfrac{1}{6}\vect{C}_{r}(\vect{z},\vect{z},\vect{z}),
\label{eq:reduced}
\end{equation}
with $\vect{\Lambda}$ the diagonal of retained eigenvalues and the reduced tensors formed
matrix-free by directional finite differences of $\vect{F}$ along the basis vectors, so that
nothing of third order is ever assembled in the full space. The quadratic term vanishes for a
cubic spring, and it is worth noting that it vanishes here too although the flow equations are
not linear. The section is symmetric and sits at zero incidence, so the flow responds to a
pitch or plunge perturbation oddly and the even part of that response has no projection on the
retained modes. Measured over the builds of this study the largest entry of $\vect{B}_r$ is
below $10^{-5}$ of the largest entry of $\vect{C}_r$, which is the noise of the finite
differences. A purely real retained mode enters the reconstruction once, and the solver
halves its left vector to that end. The reduced model is built at a specific freestream speed.

On strip theory the eigenvectors come from a dense solve of fifteen states. On the flow model
they come from shift-invert Arnoldi iterations about a few shifts on the imaginary axis at the
uncoupled structural frequencies, each with one sparse factorisation of the shifted Jacobian
held at a time, and the left eigenvectors from the same solve on the transpose, paired to the
right ones by eigenvalue. Among the modes found, those that carry the structure are separated
from the flow's convected modes by the fraction of the eigenvector's weighted norm in the
structural entries, and their mechanism by the share of each degree of freedom in that
fraction. The basis of the companion paper is retained, two complex pairs and four real
states, the flutter pair and the second pair dominated by pitch and plunge, taken here as the
two least damped structural pairs that are not flap-dominated and that oscillate at more than
a third of the flutter frequency, the last condition excluding the flow's lag modes, which
are real or nearly so. The flap pair, the electrical mode and the real plunge lag mode of the
flow are added in turn where the basis is examined. Table~\ref{tab:modes} lists the least
damped structural modes of the flow model at the construction point of the accuracy study,
$\Us = 7.0$, plunge mounting, $\chi = 0.20$, $r = 1.0$. Beside the flutter pair and the flap
pair, both close to their strip-theory counterparts, the flow's second pitch-plunge pair sits
at a higher frequency than strip theory's, and the flow supplies a real plunge lag mode and a
sequence of pairs of mixed pitch and plunge with growing damping where strip theory has the
two Wagner exponents.

\begin{table}[H]
\centering
\caption{Least damped structural modes of the flow model at $\Us = 7.0$, plunge mounting,
$\chi = 0.20$, $r = 1.0$, $M = 0.10$, per unit $\tau$, and the transducer pole.}
\label{tab:modes}
\small
\begin{tabular}{llll}
\toprule
eigenvalue $\lambda$ & type & share of each freedom in the structural part & retained \\
\midrule
$+0.02162 \pm 0.07490\mathrm{i}$ & pair & plunge 0.91, pitch 0.09, flap 0.00 & yes \\
$-0.00717 \pm 0.56697\mathrm{i}$ & pair & plunge 0.00, pitch 0.00, flap 1.00 & no \\
$-0.02904$ & real & plunge 1.00, pitch 0.00, flap 0.00 & no \\
$-0.03969 \pm 0.08788\mathrm{i}$ & pair & plunge 0.75, pitch 0.24, flap 0.00 & yes \\
$-0.06270 \pm 0.14394\mathrm{i}$ & pair & plunge 0.20, pitch 0.75, flap 0.04 & no \\
$-0.08959 \pm 0.09722\mathrm{i}$ & pair & plunge 0.64, pitch 0.36, flap 0.01 & no \\
$-0.09712 \pm 0.18910\mathrm{i}$ & pair & plunge 0.25, pitch 0.65, flap 0.10 & no \\
$-0.11458 \pm 0.13155\mathrm{i}$ & pair & plunge 0.47, pitch 0.50, flap 0.03 & no \\
$-0.14344$ & real & transducer pole, $-1/(\Us r) = -0.14286$ & no \\
\bottomrule
\end{tabular}

\end{table}

\section{Results}
\label{sec:results}

\subsection{Verification}

The transducer enters the flow model through the same row and column of the coupled system
as in strip theory, and three checks confirm it on the paper mesh at $\Us = 7.0$, $r = 1.0$.
With the harvester enabled but $\chi = 0$ the leading eigenvalue of the coupled Jacobian is
unchanged to $3.4e-16$ and the spectrum gains one real eigenvalue at $-1/(\Us r)$, recovered to
$3.3e-16$. With the coupling on, the electrical row and column of the sparse Jacobian match
central differences of the residual to round-off, since Eq.~\eqref{eq:circuit} is linear. The
power written by a march reduces to $\chi v^2/r$ to the six digits of the record. On the
$32\times10$ mesh, where the spectrum can be computed densely, the $\chi = 0$ spectrum is the
baseline spectrum plus the electrical eigenvalue, mode for mode.

Four further checks concern the limit cycle rather than the operator, and they are the ones that
decide whether the numbers of Section~\ref{sec:accuracy} mean anything. First, every amplitude
is measured over whole cycles inside the last $1000$ time units of a march, never over the
whole record. The response overshoots on its way up to the cycle, reaching $\alpha = 0.0573$
in its fourth cycle at $\Us = 7.0$ before settling at $0.05344$, so a maximum over the whole
march would report the transient. Second, halving the
time step to $0.1$ changes the cycle at $\Us = 7.0$ by less than $0.01\%$ in every quantity, so
the second-order scheme is converged in time at the step used. Third, the same cycle on the
coarser $64\times24$ mesh differs by $1.0\%$ in the mean power and $0.6\%$ in the pitch
amplitude, so the paper mesh is converged for the nonlinear response as well as for the
boundary. Fourth, the strip-theory limit cycle marched through the same front end as every
flow-model result here reproduces the companion paper's own record at $\Us = 7.0$ to better than
$0.05\%$ in the amplitudes, the voltage, the mean power and the period. The last of these
matters most, since it means that the differences reported below between the two aerodynamic
models are differences of aerodynamics and not of integrator, measurement window or
post-processing.

\subsection{Baseline flutter boundary, strip theory against the Euler model}

Without the transducer the flutter boundary of the flow model lies below the strip-theory one
at every Mach number (Table~\ref{tab:baseline}), by $-0.56\%$ at $M = 0.10$ and $-2.08\%$ at
$M = 0.30$, and the flutter frequency is within $1.3\%$ of strip theory's $0.0846$. The shift
at the lowest Mach number is the thickness of the section and the discretisation, the growth
with Mach number is compressibility. The mesh check at $M = 0.10$ spans $0.16\%$ from
$64\times24$ to $128\times48$, and the paper mesh sits within $0.08\%$ of the finest, so the
mesh contributes less to the comparison than the aerodynamic model does. A flutter boundary
costs eleven to thirteen spectra of the coupled Jacobian, found by the Illinois rule on a
bracket, four to seven minutes on the paper mesh.

\begin{table}[H]
\centering
\caption{Linear flutter boundary of the baseline section without the transducer, strip theory
and the Euler model at three Mach numbers on the paper mesh, with the mesh check at $M = 0.10$.}
\label{tab:baseline}
\small
\begin{tabular}{llrrr}
\toprule
model & $M$ & $\Us_F$ & $\omega_F$ & shift [\%] \\
\midrule
strip theory & incompressible & 6.3246 & 0.0846 & \\
Euler & 0.10 & 6.2889 & 0.0835 & $-0.56$ \\
Euler & 0.20 & 6.2659 & 0.0842 & $-0.93$ \\
Euler & 0.30 & 6.1932 & 0.0843 & $-2.08$ \\
Euler, 64$\times$24 cells & 0.10 & 6.2835 & 0.0838 & $-0.65$ \\
Euler, 128$\times$48 cells & 0.10 & 6.2939 & 0.0833 & $-0.49$ \\
\bottomrule
\end{tabular}

\end{table}

\subsection{Effect of the transducer on the flutter boundary}

\begin{figure}[H]
\centering
\includegraphics[width=\textwidth]{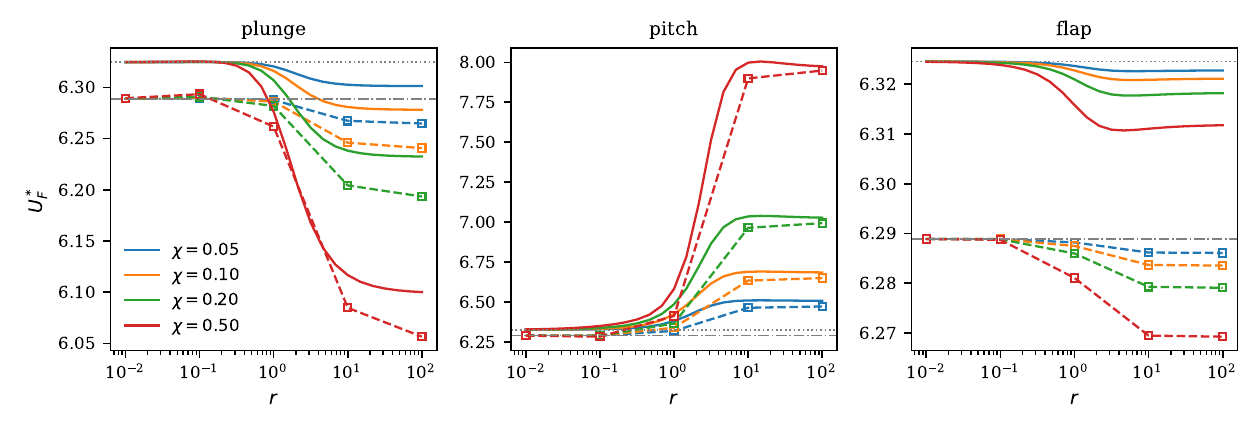}
\caption{Linear flutter boundary against load resistance, transducer on the plunge, pitch and
flap degree of freedom. Solid lines strip theory, dashed lines with markers the Euler model at
$M = 0.10$, the horizontal lines the two baselines.}
\label{fig:A}
\end{figure}

Table~\ref{tab:flutter} and Figure~\ref{fig:A} give the linear flutter boundary against load
resistance for the transducer on each degree of freedom, strip theory and the flow model at
$M = 0.10$, on the load resistances of the companion paper's table. The strip-theory values
reproduce that table to its rounding. Because the two baselines differ, the comparison is of
the shift each model predicts from its own baseline. A plunge-mounted transducer is
destabilising in both, and the shifts agree wherever they are large, at $\chi = 0.50$ by
$-3.29\%$ against $-3.24\%$ at $r = 10$ and $-3.55\%$ against $-3.69\%$ in the
open-circuit limit, strip theory first. A pitch-mounted transducer is stabilising in both,
by $+26.5\%$ against $+25.6\%$ at $\chi = 0.50$, $r = 10$, and both reach the same
open-circuit plateau. The one systematic difference is at $r = 1$, where the electrical decay
rate $\lambda = 1/(\Us r)$ is comparable to the flutter frequency and the transducer acts as a
damper rather than a stiffness. There the flow model gives about half the shift of strip
theory for the pitch mounting at every coupling, $+4.06\%$ against $+1.95\%$ at
$\chi = 0.50$, and the same tendency for the plunge mounting, $-0.76\%$ against
$-0.43\%$. A flap-mounted transducer, not considered in the companion paper, is
destabilising and an order of magnitude weaker than the plunge mounting in both models, at
most $-0.21\%$ against $-0.31\%$ at $\chi = 0.50$. Both mountings recover the baseline as
$r \to 0$, since a short-circuited transducer exerts no force, and the open-circuit limit is
an added stiffness, as in strip theory.

\begin{table}[H]
\centering
\caption{Linear flutter boundary against load resistance for the transducer on each degree
of freedom at four couplings, strip theory and the Euler model at $M = 0.10$.}
\label{tab:flutter}
\scriptsize
\begin{tabular}{llrrrrr}
\toprule
mounting & $\chi$ & \multicolumn{5}{c}{$\Us_F$, strip theory / Euler $M = 0.10$} \\
\cmidrule(lr){3-7}
 & & $r = 0.01$ & $r = 0.1$ & $r = 1$ & $r = 10$ & $r = 100$ \\
\midrule
plunge & 0.05 & 6.3246 / 6.2889 & 6.3247 / 6.2893 & 6.3204 / 6.2877 & 6.3026 / 6.2673 & 6.3013 / 6.2647 \\
plunge & 0.10 & 6.3246 / 6.2890 & 6.3247 / 6.2898 & 6.3161 / 6.2861 & 6.2808 / 6.2460 & 6.2781 / 6.2408 \\
plunge & 0.20 & 6.3246 / 6.2891 & 6.3249 / 6.2907 & 6.3069 / 6.2819 & 6.2382 / 6.2043 & 6.2325 / 6.1935 \\
plunge & 0.50 & 6.3247 / 6.2894 & 6.3252 / 6.2933 & 6.2763 / 6.2617 & 6.1168 / 6.0849 & 6.1002 / 6.0567 \\
\addlinespace
pitch & 0.05 & 6.3249 / 6.2888 & 6.3279 / 6.2888 & 6.3801 / 6.3182 & 6.5090 / 6.4630 & 6.5067 / 6.4710 \\
pitch & 0.10 & 6.3252 / 6.2888 & 6.3310 / 6.2885 & 6.4232 / 6.3384 & 6.6886 / 6.6331 & 6.6843 / 6.6490 \\
pitch & 0.20 & 6.3258 / 6.2887 & 6.3365 / 6.2875 & 6.4850 / 6.3649 & 7.0352 / 6.9639 & 7.0274 / 6.9933 \\
pitch & 0.50 & 6.3274 / 6.2883 & 6.3489 / 6.2819 & 6.5812 / 6.4116 & 7.9987 / 7.8999 & 7.9753 / 7.9491 \\
\addlinespace
flap & 0.05 & 6.3246 / 6.2889 & 6.3245 / 6.2889 & 6.3237 / 6.2882 & 6.3227 / 6.2861 & 6.3228 / 6.2861 \\
flap & 0.10 & 6.3246 / 6.2889 & 6.3245 / 6.2888 & 6.3228 / 6.2875 & 6.3209 / 6.2836 & 6.3211 / 6.2835 \\
flap & 0.20 & 6.3246 / 6.2889 & 6.3243 / 6.2888 & 6.3210 / 6.2860 & 6.3178 / 6.2792 & 6.3182 / 6.2791 \\
flap & 0.50 & 6.3245 / 6.2889 & 6.3239 / 6.2887 & 6.3158 / 6.2810 & 6.3111 / 6.2694 & 6.3118 / 6.2692 \\
\bottomrule
\end{tabular}

\end{table}

\subsection{Where to put the transducer, and how much it harvests}
\label{sec:powermap}

\begin{figure}[H]
\centering
\includegraphics[width=\textwidth]{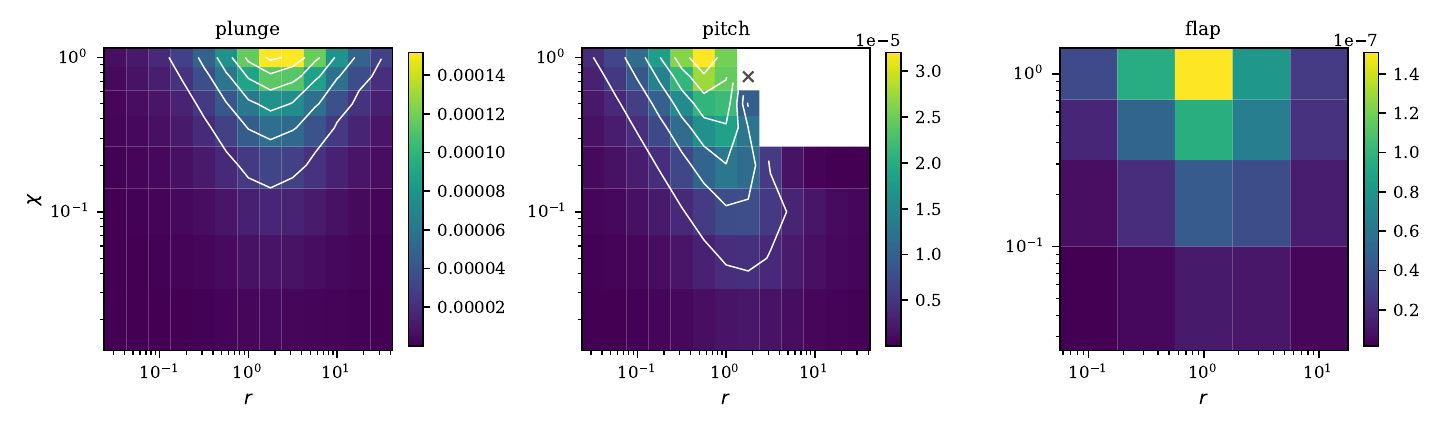}
\caption{Mean power over the coupling and load-resistance plane at $\Us = 7.0$, the flow model's
reduced model in colour and the strip-theory map on the same grid as white contours. Blank
cells are linearly stable, and the cross marks the linearly unstable cell that had not settled.}
\label{fig:B}
\end{figure}

The companion paper mapped the mean power over the coupling and load-resistance plane at
$\Us = 7.0$ and found the mounting to be a first-order design variable. Figure~\ref{fig:B} and
Table~\ref{tab:power} repeat that map on the flow model, on the companion paper's own grid of
eight couplings and thirteen load resistances for the plunge and the pitch mounting, so that the
two maps are compared cell for cell and nothing is interpolated, and on a coarser grid for the
flap mounting, which that paper did not try. There is one difference of method. A full-model
limit cycle costs about half an hour of computing per cell, so the map is computed with the
reduced model of Section~\ref{sec:accuracy} rebuilt at each cell, and six cells are recomputed
with the full model as anchors, fixing the scale.

The design findings survive the change of aerodynamic model without change. The optimum lies in
the same cell in both models for both mountings, at $\chi = 1.00$, $r = 1.78$ for the plunge
mounting and at $\chi = 1.00$, $r = 0.56$ for the pitch mounting. On the common mechanical
reference the best plunge cell of the reduced-model map reaches $15.18\times10^{-5}$ and the
best pitch cell $3.21\times10^{-5}$, a factor of $4.7$ in favour of plunge, and the full-model
anchors at the same two cells give $11.44\times10^{-5}$ and $2.96\times10^{-5}$, a factor of
$3.9$, against the factor of $4.5$ the companion paper reports. The ranking of the mountings is
therefore unchanged and its margin is about the same. The flap mounting reaches
$1.51\times10^{-7}$, three orders below the plunge mounting, so a transducer on the control
surface is not an option worth pursuing on this section.

Of the $104$ pitch-mounted cells, $83$ are linearly unstable on the flow model at this velocity
against $78$ in strip theory. The five cells that oscillate on the flow model only, four at
$\chi = 0.20$ with $r \ge 5.6$ and one at $\chi = 0.75$ with $r = 1.78$, lie just above the
pitch-mounted flutter boundary, which Table~\ref{tab:flutter} puts lower on the flow model than
in strip theory. Wherever the map and that table share a cell, $33$ cells in all, the cell has a
limit cycle if and only if its linear boundary lies below $\Us = 7.0$, which is an agreement
between two independent computations rather than a restatement of one. Four of the marginal
cells had not settled within the horizon of the map and were marched again to $12\,000$ time
units, where three settled. The fourth, at $\chi = 0.75$ and $r = 1.78$, is linearly unstable but
still growing, and is marked in Figure~\ref{fig:B} rather than given a value.

The anchors separate two things that the reduced-model map mixes. Between the two full models,
with no reduced model in either path, the flow model harvests $+22\%$ to $+39\%$ more than strip
theory at the four plunge anchors and $+23\%$ and $+42\%$ at the two pitch anchors, with no
trend in load resistance, so strip theory under-predicts the harvested power of this section by
a fifth to two fifths. The reduced model in turn over-predicts the full model by an amount that
grows with the coupling and with the load resistance, $+5.8\%$, $+32.7\%$ and $+39.1\%$ at
$r = 0.316$, $1.78$ and $10$ for $\chi = 1.00$ against $+7.5\%$ at $r = 1.78$ for
$\chi = 0.05$. Read cell by cell against the companion paper's map, the reduced-model map is
$1.38$ to $1.95$ times it for the plunge mounting, with a median of $1.43$, rising with the load
resistance, and the anchors show that the rise is the reduced model's own over-prediction and
not a property of the flow. So the map carries the position of the optimum and the ranking of
the mountings, which the anchors confirm, and the anchors carry the magnitude.

\begin{table}[H]
\centering
\caption{Mean power at $\Us = 7.0$ on the common mechanical reference at the cells recomputed with
the full model, beside the strip-theory full model and the flow model's reduced-model map in the
same cell. An asterisk marks the optimum cell of a mounting, the same in both aerodynamic models.}
\label{tab:power}
\footnotesize
\begin{tabular}{lrrrrrrr}
\toprule
 & & & \multicolumn{3}{c}{mean power $\times10^{5}$} & \multicolumn{2}{c}{difference [\%]} \\
\cmidrule(lr){4-6}\cmidrule(lr){7-8}
mounting & $\chi$ & $r$ & strip, full & flow, reduced & flow, full & full, flow to strip & flow, reduced to full \\
\midrule
plunge & 1.00 & 0.316 & $3.44$ & $4.76$ & $4.50$ & $+30.7$ & $+5.8$ \\
plunge$^\ast$ & 1.00 & 1.78 & $9.35$ & $15.18$ & $11.44$ & $+22.3$ & $+32.7$ \\
plunge & 1.00 & 10 & $4.22$ & $7.83$ & $5.63$ & $+33.5$ & $+39.1$ \\
plunge & 0.05 & 1.78 & $0.54$ & $0.80$ & $0.75$ & $+39.1$ & $+7.5$ \\
pitch$^\ast$ & 1.00 & 0.562 & $2.09$ & $3.21$ & $2.96$ & $+41.7$ & $+8.2$ \\
pitch & 0.20 & 1 & $0.89$ & $1.27$ & $1.09$ & $+22.8$ & $+16.4$ \\
flap$^\ast$ & 1.00 & 1 & -- & $0.0151$ & -- & -- & -- \\
\bottomrule
\end{tabular}

\end{table}

\subsection{Limit-cycle oscillation and harvested power}

\begin{figure}[H]
\centering
\includegraphics[width=\textwidth]{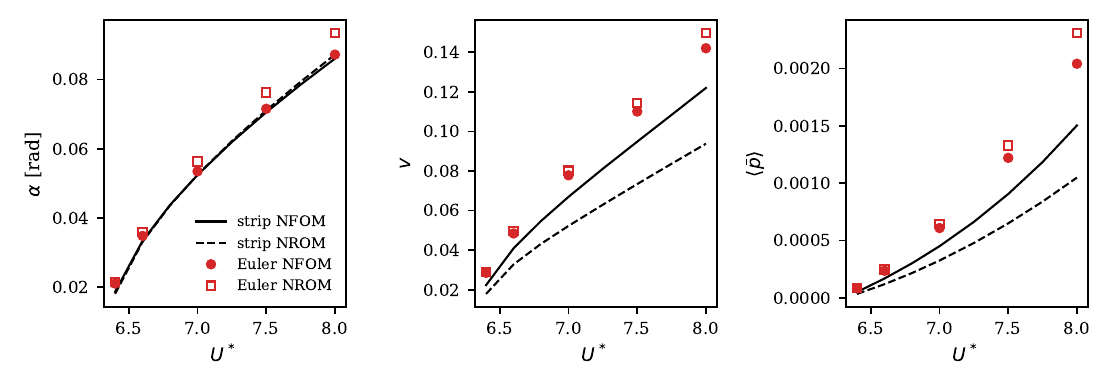}
\caption{Limit-cycle pitch amplitude, voltage and mean power against reduced velocity,
strip theory full (solid) and reduced (dashed) models from the companion paper, the flow
model's full model (filled) and reduced model (open).}
\label{fig:C}
\end{figure}

Above the flutter boundary the response of the flow model settles onto a limit cycle, as in
strip theory. Table~\ref{tab:lco} and Figure~\ref{fig:C} give the amplitudes, the voltage and
the mean power at five reduced velocities for the plunge mounting at $\chi = 0.20$, $r = 1.0$,
strip theory from the companion paper's own records and the flow model from marches of $3000$
time units, about $40$ cycles. Every amplitude here is measured over whole cycles inside the
LAST $1000$ time units, never over the whole record, because the response overshoots on its way
up, reaching $\alpha = 0.0573$ in its fourth cycle and settling at $0.05344$, so a maximum
taken over the whole march would report the transient and not the limit cycle. Inside that
window the last four cycles agree in amplitude to better than $3e-05$ in every case.

The cycle of the flow model is larger than the strip-theory one at the same reduced velocity
throughout the range, and the gap widens with velocity. The pitch amplitude is $+12.7\%$ at
$\Us = 6.4$ and $+1.4\%$ at $8.0$, the plunge amplitude $+29\%$ rising to $+18\%$, the
voltage $+28\%$ to $+16\%$ and the mean power $+63\%$ to $+36\%$, with the period
within $1.5\%$. Part of this is the position of the boundary, since at the same $\Us$ the flow
model is further above its own flutter speed, but the excess velocity accounts for a few percent
of amplitude only. The rest is the cycle itself, the balance between the cubic spring and the
aerodynamic work over a cycle, and it falls mostly on the plunge, the ratio of plunge to pitch
amplitude in the cycle being $2.89$ in the flow model against $2.52$ in strip theory at
$\Us = 7.0$. The harvested power follows the voltage, and the voltage follows the plunge
amplitude through Eq.~\eqref{eq:circuit}, so the flow model harvests about a third more than
strip theory predicts at the top of the range. The two models agree on the character of the
bifurcation, an amplitude growing as the square root of the excess velocity and a power growing
as its square.

\begin{table}[H]
\centering
\caption{Limit cycle of the plunge-mounted harvester at $\chi = 0.20$, $r = 1.0$,
$\beta_\alpha = 100$, strip theory and the flow model at $M = 0.10$, full and reduced models.
Amplitudes in radians and semi-chords, mean power on the mounting's own reference.}
\label{tab:lco}
\footnotesize
\begin{tabular}{l rrrr rrrr}
\toprule
 & \multicolumn{4}{c}{strip theory, NFOM} & \multicolumn{4}{c}{strip theory, NROM} \\
\cmidrule(lr){2-5}\cmidrule(lr){6-9}
$\Us$ & $\alpha$ & $\xi$ & $v$ & $\langle\bar p\rangle$ & $\alpha$ & $\xi$ & $v$ & $\langle\bar p\rangle$ \\
\midrule
6.20 & -- & -- & -- & -- & -- & -- & -- & -- \\
6.30 & -- & -- & -- & -- & -- & -- & -- & -- \\
6.40 & 0.0186 & 0.0461 & 0.0224 & 5.00e-05 & 0.0179 & 0.0431 & 0.0179 & 3.30e-05 \\
6.60 & 0.0333 & 0.0831 & 0.0409 & 1.68e-04 & 0.0330 & 0.0809 & 0.0328 & 1.18e-04 \\
6.80 & 0.0437 & 0.1094 & 0.0548 & 3.01e-04 & 0.0436 & 0.1087 & 0.0432 & 2.15e-04 \\
7.00 & 0.0524 & 0.1318 & 0.0670 & 4.50e-04 & 0.0524 & 0.1328 & 0.0523 & 3.24e-04 \\
7.25 & 0.0619 & 0.1571 & 0.0812 & 6.62e-04 & 0.0622 & 0.1604 & 0.0630 & 4.77e-04 \\
7.50 & 0.0705 & 0.1808 & 0.0948 & 9.05e-04 & 0.0711 & 0.1862 & 0.0734 & 6.49e-04 \\
7.75 & 0.0785 & 0.2039 & 0.1084 & 1.18e-03 & 0.0794 & 0.2110 & 0.0837 & 8.39e-04 \\
8.00 & 0.0861 & 0.2268 & 0.1220 & 1.50e-03 & 0.0872 & 0.2350 & 0.0938 & 1.05e-03 \\
\bottomrule
\end{tabular}
\\[6pt]
\begin{tabular}{l rrrr rrrr}
\toprule
 & \multicolumn{4}{c}{Euler, NFOM} & \multicolumn{4}{c}{Euler, NROM} \\
\cmidrule(lr){2-5}\cmidrule(lr){6-9}
$\Us$ & $\alpha$ & $\xi$ & $v$ & $\langle\bar p\rangle$ & $\alpha$ & $\xi$ & $v$ & $\langle\bar p\rangle$ \\
\midrule
6.20 & -- & -- & -- & -- & -- & -- & -- & -- \\
6.30 & -- & -- & -- & -- & -- & -- & -- & -- \\
6.40 & 0.0209 & 0.0594 & 0.0285 & 8.15e-05 & 0.0213 & 0.0613 & 0.0290 & 8.42e-05 \\
6.60 & 0.0347 & 0.0992 & 0.0484 & 2.35e-04 & 0.0359 & 0.1085 & 0.0497 & 2.47e-04 \\
6.80 & -- & -- & -- & -- & -- & -- & -- & -- \\
7.00 & 0.0534 & 0.1546 & 0.0778 & 6.08e-04 & 0.0563 & 0.1974 & 0.0802 & 6.44e-04 \\
7.25 & -- & -- & -- & -- & -- & -- & -- & -- \\
7.50 & 0.0716 & 0.2121 & 0.1101 & 1.22e-03 & 0.0763 & 0.3299 & 0.1145 & 1.33e-03 \\
7.75 & -- & -- & -- & -- & -- & -- & -- & -- \\
8.00 & 0.0873 & 0.2673 & 0.1420 & 2.04e-03 & 0.0934 & 0.4914 & 0.1496 & 2.31e-03 \\
\bottomrule
\end{tabular}

\end{table}

\subsection{Accuracy of the reduced models on the Euler full-order model}
\label{sec:accuracy}

The reduced model is built at each reduced velocity on the two pairs of Table~\ref{tab:modes}
with the cubic terms, marched from the same initial condition to the same horizon as the full
model and measured in the same window. Table~\ref{tab:nrom} gives its differences from the full
model across the range, and Figure~\ref{fig:C} shows them beside the strip-theory pair. The
linear reduced model grows without bound above the flutter boundary, as it must, and is not
tabulated. The harvested quantities are reproduced well and the plunge is not. The voltage stays
within $5.3\%$ of the full model and the mean power within $+13.0\%$, the period within
$1.8\%$ and the pitch amplitude within $+7.1\%$, while the plunge amplitude is $+3.3\%$ at
$\Us = 6.4$ and grows with the excess velocity to $+84\%$ at $8.0$. The flap deflection, which
no retained mode carries, is $-7\%$ to $-75\%$ low, as in strip theory. The picture is the
reverse of the strip-theory one in one respect and the same in another. There the four-state
model reproduced pitch and plunge to a percent and lost a fifth of the voltage. Here it
reproduces the voltage and the power to a few percent and loses the plunge.

\begin{table}[H]
\centering
\caption{Differences of the two-pair reduced model from the full model on the flow model,
plunge mounting, $\chi = 0.20$, $r = 1.0$, $\beta_\alpha = 100$, in percent of the full-model
value, across reduced velocity.}
\label{tab:nrom}
\small
\begin{tabular}{lrrrrrr}
\toprule
$\Us$ & $\alpha$ & $\xi$ & $\delta$ & $v$ & $\langle\bar p\rangle$ & period \\
\midrule
6.40 & $+1.9$ & $+3.3$ & $-7.5$ & $+1.7$ & $+3.3$ & $+0.1$ \\
6.60 & $+3.4$ & $+9.5$ & $-20.4$ & $+2.5$ & $+4.9$ & $+0.1$ \\
7.00 & $+5.4$ & $+27.7$ & $-40.9$ & $+3.0$ & $+6.0$ & $-0.2$ \\
7.50 & $+6.6$ & $+55.5$ & $-60.9$ & $+4.0$ & $+8.9$ & $-0.8$ \\
8.00 & $+7.1$ & $+83.9$ & $-74.7$ & $+5.3$ & $+13.0$ & $-1.8$ \\
\bottomrule
\end{tabular}

\end{table}

Table~\ref{tab:variants} examines the basis at $\Us = 7.0$ and at $8.0$. Adding the flap pair
recovers the flap deflection, from $-41\%$ to $+30\%$, and changes nothing else, so the flap
error is plain truncation, as it was in strip theory. Adding the electrical mode moves the
voltage from $+3.0\%$ to $+0.3\%$ and the power from $+6.0\%$ to $+1.5\%$. Adding the real
plunge lag mode makes the voltage $+30\%$ and the power $+76\%$ high while leaving the plunge
where it was, an effect of the same kind the gust study of this solver met, that the modes of a
non-normal system do not add up one at a time, so that a mode carrying part of the true response
can worsen the reduced one when its partners are absent.

Enlarging the basis does not remove the plunge error either, and it costs the harvested
quantity. At $\Us = 8.0$ two, four and six structural pairs give a pitch error of $+7.1\%$,
$+2.6\%$ and $+2.3\%$ and a plunge error of $+84\%$, $+47\%$ and $+44\%$, so the
improvement stops between four pairs and six, while the voltage goes from $+5.3\%$ to
$-18\%$ and $-20\%$ and the power from $+13\%$ to $-35\%$ and $-38\%$. An error that
does not vanish as the basis grows is not truncation, and a basis that improves the structural
motion while spoiling the harvested quantity is not simply too small. Section~\ref{sec:span}
separates what the basis cannot represent from what the reduced march does with what it can.

\begin{table}[H]
\centering
\caption{Differences of the reduced model from the full model at $\Us = 7.0$ on the flow
model for the bases examined, in percent of the full-model value.}
\label{tab:variants}
\small
\begin{tabular}{lrrrrrrr}
\toprule
basis & modes & $\alpha$ & $\xi$ & $\delta$ & $v$ & $\langle\bar p\rangle$ & period \\
\midrule
flutter pair and second pair & 2 & $+5.4$ & $+27.7$ & $-40.9$ & $+3.0$ & $+6.0$ & $-0.2$ \\
flutter pair and second pair ($\Us = 8$) & 2 & $+7.1$ & $+83.9$ & $-74.7$ & $+5.3$ & $+13.0$ & $-1.8$ \\
and the flap pair & 3 & $+5.4$ & $+28.4$ & $+29.6$ & $+3.6$ & $+7.3$ & $-0.2$ \\
and the electrical mode & 3 & $+5.4$ & $+27.8$ & $-40.7$ & $+0.3$ & $+1.5$ & $-0.2$ \\
and the plunge lag mode & 3 & $+6.1$ & $+29.6$ & $-39.1$ & $+30.3$ & $+75.8$ & $+1.8$ \\
and all three & 5 & $+6.0$ & $+30.6$ & $+38.1$ & $+28.2$ & $+67.3$ & $+1.8$ \\
four structural pairs & 4 & $+1.8$ & $+10.4$ & $-48.3$ & $-11.6$ & $-22.9$ & $-0.8$ \\
four structural pairs ($\Us = 8$) & 4 & $+2.6$ & $+47.0$ & $-83.2$ & $-18.3$ & $-35.1$ & $-3.3$ \\
six structural pairs ($\Us = 8$) & 6 & $+2.3$ & $+44.4$ & $-82.3$ & $-19.9$ & $-37.7$ & $-3.6$ \\
\bottomrule
\end{tabular}

\end{table}

\subsection{What the basis can represent, and what the march does with it}
\label{sec:span}

An error of a reduced model has two parts that are worth separating. The basis may be unable to
represent the true response at all, and the reduced march may not find the best answer inside
the part it can represent. The companion paper separated them for its voltage by projecting the
true limit cycle onto the retained subspace. The same is done here. The full model is marched
again at $\Us = 7.0$ with its whole state kept over the last cycles, and that state is projected
with the left eigenvectors onto each candidate basis, which gives the best the basis could
possibly do.

The two-pair basis represents the true cycle to $-11.9\%$ in pitch, $+7.0\%$ in plunge and
$-13.4\%$ in the voltage, while the reduced march delivers $+5.4\%$, $+27.7\%$ and
$+3.0\%$. Two things follow. In pitch and in the voltage the representation error and the
dynamics error have opposite signs and largely cancel, so the good agreement of
Table~\ref{tab:nrom} in those two quantities is partly a coincidence of signs and not a sign
that the basis is adequate. In plunge they add. With four pairs the representation improves to
$-3.2\%$ in pitch and $+6.4\%$ in plunge, and the reduced march improves with it, which is
the ordinary behaviour of a truncated basis.

\subsection{The dynamic response, and the shape of the cycle}
\label{sec:shape}

\begin{figure}[H]
\centering
\includegraphics[width=\textwidth]{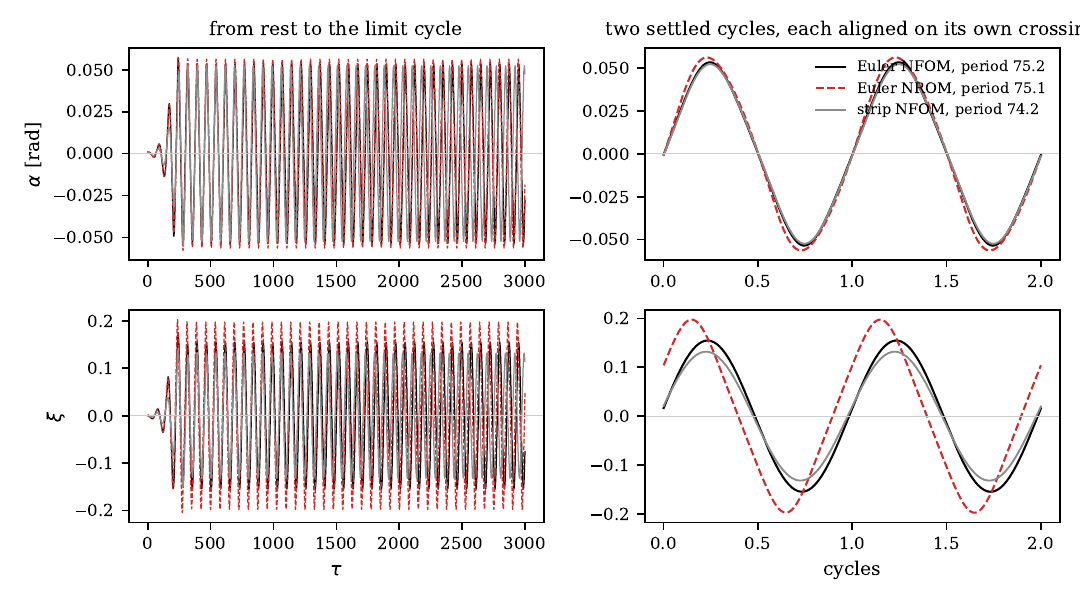}
\caption{Pitch and plunge at $\Us = 7.0$, plunge mounting, $\chi = 0.20$, $r = 1.0$. Left, from
rest to the limit cycle. Right, two settled cycles with each model aligned on its own upward
crossing.}
\label{fig:D}
\end{figure}

Figure~\ref{fig:D} shows the response itself, pitch and plunge, from rest to the limit cycle and
then over two settled cycles with each model aligned on its own zero crossing, since the periods
differ and the absolute phase after forty cycles carries no information. The pitch traces of the
three models lie almost on top of one another. The plunge traces do not, and the difference is
not only one of size. In the settled cycle of the full model the plunge amplitude is $2.893$
times the pitch amplitude and the plunge lags the pitch by $6.0$ degrees. In the reduced model
the ratio is $3.507$ and the lag is $35.9$ degrees. The reduced model is therefore not
reproducing the same orbit with a slightly larger plunge. It is tracing a differently shaped
orbit.

That shape is a property of the linear subspace and not of the nonlinearity, which three
measurements establish together. First, the shape of the true cycle does not depend on its
size. As the cubic spring is changed to make the cycle four times larger, the measured plunge to pitch
ratio of the full model stays at $2.892$, $2.893$ and $2.899$. Second, the error of the reduced
model does not depend on it either, the plunge error moving only from $+26.8\%$ to $+31.3\%$
across that fourfold range, when a shortcoming of the cubic expansion of a nonlinear flow
residual would fall roughly with the square of the amplitude. Third, the retained flutter
eigenvector carries a plunge to pitch ratio of $3.173$ at a phase of $+47.3$ degrees, which is
neither the true cycle's shape nor quite the reduced model's, so the true cycle draws its shape
from content the retained pair does not hold, and the reduced model relaxes towards the shape
its own eigenvector prescribes.

This is the substantive difference between the two aerodynamic models. In strip theory the
aerodynamic states are eight lag variables attached to the structural motion, and two pairs span
the shape of the cycle almost exactly, which is why the companion paper found its structural
response accurate to a percent and its error concentrated in the transducer. In a compressible
flow the aerodynamic state is a field, the cycle's shape is set partly by flow content spread
across the continuum, and a basis of a few coupled eigenvectors captures the frequency and the
pitch amplitude of that cycle while getting the proportion of plunge to pitch wrong by a fifth.
A reduced model of this kind, on this class of aerodynamics, should therefore be reported on the
shape of the orbit and not on one amplitude.

\section{Discussion}
\label{sec:discussion}

The study finds that the degree of freedom that carries the transducer still sets the sign of the shift in the flutter boundary, a plunge
mounting lowering it and a pitch mounting raising it, and the shifts agree with strip theory
wherever they are large. It still sets the harvested power, by about the same margin, a factor
of $3.9$ between the plunge and the pitch mounting on the full model and $4.7$ on the reduced one
against $4.5$ in strip theory, with the optimum in the same cell of the coupling and load plane
and the flap mounting three orders below either. The one systematic disagreement is at a load
resistance near unity, where the electrical time constant is comparable with the flutter
frequency and the transducer acts as a damper rather than as a stiffness. There the flow model
gives about half the shift strip theory predicts. A designer using strip theory to choose a
mounting would choose correctly, and one using it to size the load near that point would not.

The magnitude of the harvested power is under-predicted by strip theory, by $+22\%$ at the
plunge optimum and by up to $+42\%$ at a pitch cell, measured between the two full models with no
reduced model in either path. The reason is visible in the limit cycle rather than in the
boundary. At the same reduced velocity the compressible section sustains a larger orbit, and
since the voltage follows the plunge velocity through Eq.~\eqref{eq:circuit} and the power
follows the voltage squared, a moderate difference in the orbit becomes a large difference in
the power.

A reduced model can be built, and it behaves differently from its strip-theory counterpart. Four
states reproduce the frequency of the cycle to under two percent and its pitch amplitude to a
few percent across the range, and they reproduce the harvested voltage and power to within
$5.3\%$ and $+13.0\%$, which is better than the companion paper managed for its voltage. What they
do not reproduce is the proportion of plunge to pitch in the orbit, and that error is neither a
matter of expanding the residual to a higher order nor, principally, of retaining more modes. It
is a property of the subspace. This suggests that the accuracy of a harvester reduced model
should be reported on the shape of the orbit, the ratio and the phase of the freedoms that drive
the transducer, and not on a single amplitude, since an amplitude can agree by a cancellation of
two errors that a shape cannot.

The practical consequence for a designer is narrow but clear. A four-state model built at the
velocity of interest is an adequate tool for choosing a mounting, for placing the optimum in the
coupling and load plane, and for estimating the power to within about a tenth at weak coupling
or low load resistance, the over-prediction growing to a third at the optimum. It is not an
adequate tool for predicting the motion of the section, and a control law designed on its plunge
channel would be designed on an orbit that is a fifth too wide and thirty degrees out of phase.

\section{Conclusions}
\label{sec:conclusions}

\begin{enumerate}
\item A piezoelectric harvester was coupled to a compressible Euler
model of a three degree of freedom aerofoil, giving a full-order model of about twelve thousand
states. The strip-theory results are reproduced through the same front end to better than
$0.05\%$, so the differences reported here are differences of aerodynamics.
\item The transducer shifts the flutter boundary in the same direction and by nearly the same
amount in both aerodynamic models wherever the shift is large. Near a load resistance of unity
the flow model gives about half the shift of strip theory.
\item The compressible section sustains a larger limit cycle than strip theory at the same
reduced velocity, and harvests $+22\%$ to $+42\%$ more power at the cells the companion paper calls
optimal. Strip theory is conservative for this section.
\item The plunge mounting remains the right choice, by a factor of $3.9$ over the pitch mounting
between the full models on the common mechanical reference against $4.5$ in strip theory, with the
optimum in the same cell of the coupling and load plane in both models. A transducer on the trailing-edge flap harvests three orders
of magnitude less and is not worth pursuing.
\item A four-state nonlinear reduced model built at the velocity of interest reproduces the
frequency, the pitch amplitude, the voltage and the power of the cycle to within about a tenth,
and misstates the proportion of plunge to pitch in the orbit by a fifth and its phase by thirty
degrees. The error is amplitude-independent and survives enlargement of the basis, so it belongs
to the retained subspace and not to the order of the expansion.
\item The accuracy of a reduced model for energy harvesting in a compressible flow should
therefore be reported on the shape of the orbit as well as on the harvested quantity, since
agreement in an amplitude can be the cancellation of a representation error against a dynamics
error of opposite sign, which the projection of the true cycle onto the basis makes visible.
\end{enumerate}

\section*{Data availability}
The data, the scripts and the solver source are available from the corresponding author on
request.


\begin{thebibliography}{99}
\setlength{\itemsep}{1pt}
\small

\bibitem{erturk2010}
A.~Erturk, W.~G.~R.~Vieira, C.~De~Marqui, and D.~J.~Inman.
On the energy harvesting potential of piezoaeroelastic systems.
\emph{Applied Physics Letters}, 96(18):184103, 2010. \doi{10.1063/1.3427405}

\bibitem{dunnmon2011}
J.~A.~Dunnmon, S.~C.~Stanton, B.~P.~Mann, and E.~H.~Dowell.
Power extraction from aeroelastic limit cycle oscillations.
\emph{Journal of Fluids and Structures}, 27(8):1182--1198, 2011.
\doi{10.1016/j.jfluidstructs.2011.02.003}

\bibitem{sousa2011}
V.~C.~Sousa, M.~de~M.~Anic\'{e}zio, C.~De~Marqui, and A.~Erturk.
Enhanced aeroelastic energy harvesting by exploiting combined nonlinearities: theory
and experiment. \emph{Smart Materials and Structures}, 20(9):094007, 2011.
\doi{10.1088/0964-1726/20/9/094007}

\bibitem{abdelkefi2011}
A.~Abdelkefi, A.~H.~Nayfeh, and M.~R.~Hajj.
Modeling and analysis of piezoaeroelastic energy harvesters.
\emph{Nonlinear Dynamics}, 67(2):925--939, 2012. \doi{10.1007/s11071-011-0035-1}

\bibitem{demarqui2013}
C.~De~Marqui and A.~Erturk.
Electroaeroelastic analysis of airfoil-based wind energy harvesting using
piezoelectric transduction and electromagnetic induction.
\emph{Journal of Intelligent Material Systems and Structures}, 24(7):846--854, 2013.
\doi{10.1177/1045389X12461073}

\bibitem{bryant2011}
M.~Bryant and E.~Garcia.
Modeling and testing of a novel aeroelastic flutter energy harvester.
\emph{Journal of Vibration and Acoustics}, 133(1):011010, 2011.
\doi{10.1115/1.4002788}

\bibitem{adhikari2009}
S.~Adhikari, M.~I.~Friswell, and D.~J.~Inman.
Piezoelectric energy harvesting from broadband random vibrations.
\emph{Smart Materials and Structures}, 18(11):115005, 2009.
\doi{10.1088/0964-1726/18/11/115005}

\bibitem{abdelkefi2016}
A.~Abdelkefi.
Aeroelastic energy harvesting: a review.
\emph{International Journal of Engineering Science}, 100:112--135, 2016.
\doi{10.1016/j.ijengsci.2015.10.006}

\bibitem{daronch2013gust}
A.~Da~Ronch, N.~D.~Tantaroudas, S.~Timme, and K.~J.~Badcock.
Model reduction for linear and nonlinear gust loads analysis.
In \emph{54th AIAA/ASME/ASCE/AHS/ASC Structures, Structural Dynamics, and
Materials Conference}, Boston, MA, 8--11 April 2013. AIAA Paper 2013-1492.
\doi{10.2514/6.2013-1492}

\bibitem{daronch2013control}
A.~Da~Ronch, N.~D.~Tantaroudas, and K.~J.~Badcock.
Reduction of nonlinear models for control applications.
In \emph{54th AIAA/ASME/ASCE/AHS/ASC Structures, Structural Dynamics, and
Materials Conference}, Boston, MA, 8--11 April 2013. AIAA Paper 2013-1491.
\doi{10.2514/6.2013-1491}

\bibitem{tantaroudas2014adaptive}
N.~D.~Tantaroudas, A.~Da~Ronch, G.~Gai, K.~J.~Badcock, and R.~Palacios.
An adaptive aeroelastic control approach using non linear reduced order models.
In \emph{14th AIAA Aviation Technology, Integration, and Operations Conference},
Atlanta, GA, June 2014. AIAA Paper 2014-2590. \doi{10.2514/6.2014-2590}

\bibitem{daronch2014flutter}
A.~Da~Ronch, N.~D.~Tantaroudas, S.~Jiffri, and J.~E.~Mottershead.
A nonlinear controller for flutter suppression: from simulation to wind tunnel
testing. In \emph{55th AIAA/ASME/ASCE/AHS/ASC Structures, Structural Dynamics, and
Materials Conference}, National Harbor, MD, 13--17 January 2014. AIAA Paper
2014-0345. \doi{10.2514/6.2014-0345}

\bibitem{papatheou2013}
E.~Papatheou, N.~D.~Tantaroudas, A.~Da~Ronch, J.~E.~Cooper, and J.~E.~Mottershead.
Active control for flutter suppression: an experimental investigation.
In \emph{International Forum on Aeroelasticity and Structural Dynamics (IFASD
2013)}, Bristol, UK, 24--26 June 2013, pp.~507--520. IFASD Paper 2013-8D.

\bibitem{tantaroudas2015vfa}
N.~D.~Tantaroudas, A.~Da~Ronch, K.~J.~Badcock, Y.~Wang, and R.~Palacios.
Model order reduction for control design of flexible free-flying aircraft.
In \emph{AIAA Atmospheric Flight Mechanics Conference}, Kissimmee, FL, January 2015.
AIAA Paper 2015-0240. \doi{10.2514/6.2015-0240}

\bibitem{tantaroudas2017vfa}
N.~D.~Tantaroudas and A.~Da~Ronch.
Nonlinear reduced-order aeroservoelastic analysis of very flexible aircraft.
In P.~Marqu\'{e}s and A.~Da~Ronch, editors, \emph{Advanced UAV Aerodynamics, Flight
Stability and Control: Novel Concepts, Theory and Applications}, pp.~143--179. John
Wiley \& Sons, Chichester, 2017. \doi{10.1002/9781118928691.ch4}

\bibitem{fichera2014}
S.~Fichera, S.~Jiffri, X.~Wei, A.~Da~Ronch, N.~D.~Tantaroudas, and J.~E.~Mottershead.
Experimental and numerical study of nonlinear dynamic behaviour of an aerofoil.
In \emph{Proceedings of ISMA2014 International Conference on Noise and Vibration
Engineering}, Leuven, Belgium, 15--17 September 2014, pp.~3609--3618.

\bibitem{tantaroudas2026coupled}
N.~D.~Tantaroudas and I.~Karachalios.
Nonlinear model order reduction for coupled aeroelastic-flight dynamic systems.
arXiv:2603.15296, 2026. \doi{10.48550/arXiv.2603.15296}

\bibitem{tantaroudas2026gust}
N.~D.~Tantaroudas and I.~Karachalios.
Rapid worst-case gust identification for very flexible aircraft using reduced-order
models. arXiv:2603.16212, 2026. \doi{10.48550/arXiv.2603.16212}

\bibitem{tantaroudas2026hinf}
N.~D.~Tantaroudas and I.~Karachalios.
$H_\infty$ robust control for gust load alleviation of geometrically nonlinear
flexible aircraft. \emph{Applied and Computational Mechanics}, 2026. Article in
press, published online 7 August 2026. \doi{10.24132/acm.2026.1114}

\bibitem{daronch2014aero}
A.~Da~Ronch, A.~J.~McCracken, N.~D.~Tantaroudas, K.~J.~Badcock, H.~Hesse, and
R.~Palacios.
Assessing the impact of aerodynamic modelling on manoeuvring aircraft.
In \emph{AIAA Atmospheric Flight Mechanics Conference}, National Harbor, MD,
13--17 January 2014. AIAA Paper 2014-0732. \doi{10.2514/6.2014-0732}

\bibitem{lee1999}
B.~H.~K.~Lee, S.~J.~Price, and Y.~S.~Wong.
Nonlinear aeroelastic analysis of airfoils: bifurcation and chaos.
\emph{Progress in Aerospace Sciences}, 35(3):205--334, 1999.
\doi{10.1016/S0376-0421(98)00015-3}

\bibitem{tantaroudas2026harvest}
N.~D.~Tantaroudas, I.~Karachalios, and A.~J.~McCracken.
Transducer placement and the limits of a four-state reduced model in post-flutter
piezoelectric energy harvesting from a pitch-plunge-flap aerofoil.
arXiv:2609.12788, 2026. \doi{10.48550/arXiv.2609.12788}

\bibitem{tantaroudas2015phd}
N.~D.~Tantaroudas.
\emph{Nonlinear model order reduction and control of very flexible aircraft}.
PhD thesis, University of Liverpool, 2015.
\url{https://livrepository.liverpool.ac.uk/2012780/}

\end{thebibliography}
\end{document}